\documentclass[twocolumn,prb,showpacs,showkeys,preprintnumbers,amsmath,amssymb]{revtex4}

\usepackage[normalem]{ulem}

\usepackage{graphicx}
\usepackage{subfigure}
\usepackage{xcolor}

\usepackage{amsmath}
\usepackage{amssymb}
\usepackage{amsthm}
\usepackage[colorlinks, citecolor=blue, urlcolor=blue, bookmarks=false,
linkcolor=blue, hypertexnames=true]{hyperref}
\usepackage{hypernat}
\usepackage{placeins}
\usepackage[outdir=./graphics]{epstopdf}
\newcommand{\bra}[1]{\langle #1 \mid}
\newcommand{\ket}[1]{\mid #1 \rangle}
\newcommand{\sandwhich}[2]{\langle #1 \mid #2 \rangle}
\newcommand{\RR}{\mathbf{R}}
\newcommand{\rr}{\mathbf{r}}
\newcommand{\kk}{\mathbf{k}}
\newcommand{\ii}{\mathrm{i}}
\newcommand{\ee}{e}
\newcommand{\GG}{\mathbf{G}}
\newcommand{\qq}{\mathbf{q}}

\newcommand{\pc}{c^{\vphantom{\dagger}}}

\newcommand{\pw}{w^{\vphantom{*}}}

\newcommand{\bn}{\bar{n}}
\newcommand{\balpha}{\bar\alpha}
\newcommand{\bbeta}{\bar\beta}
\newcommand{\bgamma}{\bar\gamma}
\newcommand{\bdelta}{\bar\delta}
\newcommand{\sais}{\ifmmode\mbox{\c{S}}\else\c{S}\fi{}a\ifmmode\mbox{\c{s}}
\else\c{s}\fi{}\ifmmode \imath \else \i \fi{}o\ifmmode \breve{g}\else
\u{g}\fi{}lu~}

\begin{document}

\title{Constrained Random Phase Approximation: the spectral method}

\author{Merzuk Kaltak}
\affiliation{VASP Software GmbH, Berggasse 21/14, 1090 Vienna, Austria}
\author{Alexander Hampel}
\affiliation{VASP Software GmbH, Berggasse 21/14, 1090 Vienna, Austria}
\author{Martin Schlipf}
\affiliation{VASP Software GmbH, Berggasse 21/14, 1090 Vienna, Austria}
\author{Indukuru Ramesh Reddy}
\affiliation{Department of Physics, Kyungpook National University, Daegu, 41566,
	Korea}
\author{Bongjae Kim}
\affiliation{Department of Physics, Kyungpook National University, Daegu, 41566,
	Korea}
\author{Georg Kresse}
\affiliation{VASP Software GmbH, Berggasse 21/14, 1090 Vienna, Austria}
\affiliation{Faculty of Physics and Center for Computational Materials Science,
	University of Vienna, Kolingasse 14-16, A-1090 Vienna, Austria}
\date{\today}

\begin{abstract}
	We present a constrained Random Phase Approximation (cRPA) method, termed
	spectral cRPA (s-cRPA), and compare it to established cRPA approaches for
	Scandium and Copper by varying the 3d shell filling. The s-cRPA method
	generally produces larger Hubbard U interaction values compared to conventional
	approaches. When applied to the realistic system CaFeO$_3$, s-cRPA yields
	interaction parameters that align more closely with those required within DFT+U
	to reproduce the experimentally observed insulating state, addressing the
	metallic behaviour predicted by standard density functionals. We examine the
	issue of negative interaction values encountered in the projector cRPA method
	for filled d-shells. We show that s-cRPA provides improved numerical stability
	by preserving electron number conservation, a constraint that is violated in
	the projector cRPA method. The s-cRPA approach addresses some limitations of
	standard cRPA methods, particularly the tendency to underestimate U values,
	suggesting its potential utility for the community.  Additionally, we have
	enhanced our implementation to include computation of multi-centre interactions
	for analysing spatial decay and developed an efficient low-scaling variant
	employing a compressed Matsubara grid to obtain full frequency-dependent
	interactions.
\end{abstract}

\pacs{71.10.-w,71.15.Mb, 71.20.Nr}
\keywords{GW, DFT, PAW, RPA, c-RPA, cRPA, non-local RPA, p-cRPA, w-cRPA,
Matsubara grid}

\maketitle

\section{Introduction}
First principles methods are a cornerstone of modern theoretical physics and
chemistry. Efficient first-principles methods often rely on conventional
mean-field theories, such as Hartree-Fock or density functional theory (DFT) to
describe material properties. However, for systems with partially filled narrow
bands, i.e.  systems with localised electrons in partially filled orbitals,
these methods struggle to accurately describe the electron correlation effects
beyond the static mean-field level. Such systems are often referred to as
strongly correlated, and methods beyond static mean-field theory are necessary
to accurately describe their physical properties.

In cases where the Hartree-Fock method and DFT fall short, model Hamiltonians
that focus on a limited number of electrons near the Fermi level--- so-called
low-energy models ---attempt to effectively capture the essential physics of
strongly correlated matter. The application of model Hamiltonians to condensed
matter physics began in the 1930s with the work of Slater and
Bethe,\cite{Slater1930,Bethe1931} followed by Mott's
contributions.\cite{Mott1949} Building upon Mott's ideas, Hubbard developed an
approach in 1963\cite{Hubbard1963} that continues to be extensively studied in
the solid-state community.

The Hubbard model comprises two key components: a hopping term $t$, which
describes the probability of strongly correlated electrons moving between sites,
and a constant interaction term $U$, limited to interactions on the same site.
Despite its apparent simplicity compared to the full many-body
Hamiltonian, the Hubbard model successfully describes various collective quantum
phenomena, including the Mott transition, ferromagnetism, and the Kondo
effect.\cite{Mott1964,Mott1968,PhysRev.147.392,PhysRev.137.A1726}

While an analytical solution for the Hubbard model with one electron in one
spatial dimension can be obtained using the Bethe
ansatz,\cite{PhysRevLett.45.379,Bethe1931} explicit solutions for two or three
dimensions remain elusive. The complexity of the model increases exponentially
with the number of electrons, necessitating the use of sophisticated
approximations to find numerical solutions with reasonable computational effort.

A significant advancement in this field occurred over two decades ago with the
development of dynamical mean-field theory (DMFT), which considers the Hubbard
model in the limit of infinite dimensions.\cite{RevModPhys.68.13} DMFT is
formulated in the Green's function formalism, where instead of explicitly
calculating the many-body wavefunction, the interacting single particle Green's
function is determined, which is directly related to photo-emission
spectroscopy. Metzner and Vollhardt demonstrated that in this limit, all
contributions to the interacting Green's function vanish except for local
Feynman diagrams, substantially reducing the solution complexity of the Hubbard
model.\cite{PhysRevLett.62.324} Subsequently, Kotliar and Georges showed that
this model could be mapped onto an Anderson impurity model,\cite{PhysRev.124.41}
enabling the use of powerful quantum Monte Carlo techniques for numerical
solutions.\cite{PhysRevB.45.6479, PhysRevLett.69.1240, PhysRevLett.97.076405,
PhysRevB.76.205120, PhysRevE.87.053305, PhysRevB.86.155158}

These breakthroughs led to further investigations in the field and the
development of methods that incorporate advances in density
functional theory, such as DFT+DMFT.\cite{PSSB:PSSB200642053,
Held_DMFT+LDA2008}
DFT+DMFT combines the model aspects of DMFT with DFT, by constructing localised
orbitals of a subset of Kohn-Sham (KS) orbitals for which the
DMFT equations are solved. This combined method can also be derived formally
from a general functional.~\cite{RevModPhys.78.865}

Although the predictive power of DMFT is
well-established,\cite{PhysRevLett.87.276403, PhysRevLett.87.067205,
1367-2630-7-1-188, PhysRevB.88.165119} making the theory truly parameter-free or
{\em ab initio} remains challenging. One particularly challenging aspect of DMFT
is the choice of the effective one-centre interaction, known as {\em Hubbard}
$U$. For an isolated atom the Coulomb integrals are rather straightforward to
compute. However, in a real solid the electrons of the constructed low-energy
model interact with a much reduced Coulomb interaction that is screened by all
other electrons in the system. This $U$ is therefore a screened
Coulomb interaction and is distinguished by the bare Coulomb interaction
calculated from the overlap of the orbitals describing the model. 

Cococcioni et al. introduced a linear response method for calculating the
Hubbard $U$ parameter from first principles, rather than fitting $U$ empirically
to experimental data.\cite{DFPTUCoOCeltech,DFPTUeph,DFPTUNiOHardy} The method is
consistent with the logic of DFT+$U$, but can lead to double counting issues in
subsequent many-body treatments or to an overestimation for $U$ for filled
shells.\cite{carta2025}

The development of the constrained random phase approximation (cRPA) marked
another significant advancement in this field.\cite{PhysRevLett.90.086402,
PhysRevB.70.195104, PhysRevB.74.125106} cRPA divides the electronic system into
two subsystems: one describing the correlated degrees of freedom dominant at the
Fermi level (target or correlated space $\mathcal{C}$), and the rest comprising
the remaining Fock space, which acts as an effective medium screening the
interaction between strongly correlated electrons near the Fermi surface.

In cRPA, all polarisation effects between correlated electrons in the target
space are eliminated to avoid double-counting perturbative contributions in
subsequent model calculations. When the target states form an isolated manifold
around the Fermi surface, this removal can be performed either directly in the
plane wave basis\cite{PhysRevB.74.125106} or by solving a matrix equation in the
Wannier domain.\cite{PhysRevB.82.045105, PhysRevB.86.085117}

However, more commonly, the target states are strongly entangled with
non-correlated (s- or p-) states of the system, complicating the removal of
screening effects in the correlated subspace. In such cases, careful separation
of the target subspace $\mathcal{C}$ from the rest of the system is
necessary. Three cRPA schemes have been proposed to address this challenge.  (i)
The first method, known as disentangled cRPA\cite{PhysRevB.80.155134}, {\em
disentangles} the states around the Fermi level, enabling a simpler removal of
correlated polarisations  in the Bloch domain.\cite{PhysRevB.77.085122} This
approach requires an energy window to project the Bloch functions around the
Fermi surface onto a minimal basis set, resulting in effective interactions that
depend on this choice.\cite{PhysRevB.77.085122} (ii) The second approach, known
as weighted cRPA (w-cRPA) suggests including additional non-correlated
(non-target) states in the Wannier basis to achieve a better representation of
the Bloch states.\cite{PhysRevB.83.121101, PhysRevB.85.045132} Thus this
approach aims at conserving the band structure.  The effective interaction is
calculated using a {\em weighted} polarisability, where the weights represent
the probabilities of Bloch states being correlated.  This approach has been
developed by \sais et al.\cite{ PhysRevB.89.125110, PhysRevB.83.121101} and
implemented in various code packages.\cite{ Nakamura2021, fleurWeb,
PhysRevB.85.045132, PhysRevB.87.165118, PhysRevB.86.165105, KaltakThesis} (iii)
The third scheme, termed as projected cRPA (p-cRPA) method, can be derived from
Kubo's formalism\cite{KaltakThesis} and has been applied in literature
successfully as well.\cite{Hampel2019, Hampel2021, Muechler2022, Hampel2021-2,
Hampel2022, Hampel2023, PhysRevB.104.045134, Merkel2024, DiSante2023, Reddy2024}
Typically, p-cRPA yields larger effective interactions compared to w-cRPA,
because more screening effects are removed.

In the present work, we present a fourth method to calculate the partially
screened Coulomb interaction for entangled target spaces.  We will first
recapitulate and analyse the similarities and differences between w-cRPA and
p-cRPA, and then present a new scheme that remedies most of the shortcomings of
the previously devised schemes. We call this method the spectral cRPA (s-cRPA)
method and present its details in \ref{sec:spectral}

The paper is structured as follows. Section \ref{sec:theory} provides a brief
overview of existing cRPA methods, where we give a recap of the details of
w-cRPA and p-cRPA  in \ref{sec:weighted} and \ref{sec:projector}, respectively.
In Section \ref{sec:results}, we compare the different cRPA schemes for fcc Sc
and Cu, examining how the effective screened interaction evolves with the
filling of the d shell. We then analyse the spatial decay and frequency
dependence of the Hubbard $U$, and present results for the realistic compound
CaFeO$_3$.

\section{Theory}\label{sec:theory}

\subsection{Wannier functions and Hubbard model}

Since the Hubbard model is based on the assumption of a local Coulomb
interaction, a basis transformation of the delocalised Bloch states has to be
performed  to construct a low-energy model suitable to be solved by methods like
DMFT. To this end, Wannier functions are typically leveraged. Wannier states are
related to Bloch states through a combination of a unitary rotation $T$
and a discrete Fourier transformation with respect to
k-points\cite{PhysRevB.7.4388}
\begin{equation}
\label{eq:transformation}
\ket{w_{\alpha\RR}}=\frac{1}{N_{\kk}}\sum\limits_{n\kk}\ee^{\ii\kk\RR}T_{\alpha
n}^{(\kk)}\ket{\phi_{n\kk}}.
\end{equation}
It will be convenient to work with the mixed basis representations, also known as
analytic {\em quasi-Bloch} states\cite{PhysRevLett.98.046402}
\begin{equation}
\label{eq:mixed}
\ket{w_{\alpha\kk}}=\sum\limits_{n}T_{\alpha n}^{(\kk)}\ket{\phi_{n\kk}}.
\end{equation}
It is important to note the arbitrary nature of the Wannier transform,
since for each Bloch state the phase factors can be arbitrarily
chosen.\cite{PhysRevB.7.4388, RevModPhys.84.1419} Marzari and Vanderbilt
proposed a maximally localised scheme, which chooses the phases to minimise a
spread functional to produce maximally localised Wannier functions
(MLWFs).\cite{PhysRevB.56.12847, PhysRevB.81.054434} This approach aims to
maximise the spatial localisation of the Wannier functions. However, in
practical applications often so-called {\em first guess}\cite{Mostofi2008685}
Wannier functions are employed, in which the rotation in
Eq.(\ref{eq:transformation}) is defined as:
\begin{equation}
\label{eq:guess}
T_{\alpha n}^{(\kk)}=\sandwhich{\phi_{n\kk}}{Y_\alpha},
\end{equation}
where $|Y_\alpha\rangle$ indicates {\rm e.g.} eigenfunctions of the hydrogen
atom. An additional orthonormalisation of the resulting orbitals is commonly
performed.  The preference for Eq. (\ref{eq:guess}) is due to its tendency to
produce Wannier functions that respect symmetries of the system, whereas MLWFs
often break the symmetry. In case of entangled bands (crossing bands in the
Brillouin zone), another crucial step is the disentanglement of the Wannier
functions.  The disentanglement procedure is performed before the
orthonormalisation and maximally localisation procedures, in order to optimise
the KS states in the chosen target space and ensure that the eigenvalues vary
smoothly. In this study, we exclusively consider Wannier functions obtained from
first guesses (but including disentanglement) for easier comparability with
other studies.

The main focus of this paper is the model Hamiltonian
\begin{equation}\label{eq:ham}
H=\sum\limits_{\bar\alpha\bar\beta}t_{\bar\alpha\bar\beta}
c^\dagger_{\bar\alpha}\pc_{\bar\beta}+
\sum\limits_{\bar\alpha\bar\beta\bar\gamma\bar\delta}'U_{\bar\alpha\bar\beta\bar\gamma\bar\delta}
c^\dagger_{\bar\alpha} c^\dagger_{\bar\beta} \pc_{\bar\delta} \pc_{\bar\gamma},
\end{equation}
where the compact notation $\bar\alpha=(\alpha,{\bf R}_\alpha)$ has been
introduced and the sum in the second term is restricted to the same unit cell
$R_{\alpha}=R_{\beta}=R_\gamma=R_\delta$.  The first term
$t_{\bar\alpha\bar\beta}$ describes the electron hopping from the Wannier state
$\ket{w_{\bar\alpha}}$ to state $\ket{w_{\bar\beta}}$ between the real lattice
site $\RR_{\alpha}$ and $\RR_{\beta}$. These matrix elements are obtained by
evaluating the Kohn-Sham Hamiltonian in the Wannier basis given by the
transformation in Eq.  (\ref{eq:guess}). For further information see also
Ref.~\onlinecite{LDA+DMFT}.  It is common practice in DMFT to treat the hopping
between two different sites, as well as the Coulomb interaction between sites on
the mean field level, using for instance, density functional theory or another
simplified theory such as GW. 

\subsection{Coulomb interaction and cRPA}\label{sec:crpa}

We focus our attention on the determination of the second term
$U_{\alpha\beta\gamma\delta}$ located at $\RR=\mathbf0$. This term describes the
effective interaction between two electron-hole pairs
$\ket{\pw_\alpha,\pw_\beta},\ket{\pw_\delta,\pw_\gamma}$ and can be approximated
as the expectation value of an effectively screened Coulomb kernel
$\mathbf{U}$\cite{PhysRevB.86.165105} at zero frequency
\begin{equation}
\label{eq:matrix}
	U_{\alpha\beta\gamma\delta}=\bra{\pw_\alpha,
\pw_\beta}\mathbf{U}(\omega=0)\ket{\pw_\delta,\pw_\gamma}.
\end{equation}
In this paper, we study and require only the on-site {\em Hubbard-Kanamori}
interaction\cite{PhysRevB.86.165105}
\begin{eqnarray}
\label{eq:Udef}
U=\frac{1}{N}\sum\limits_{\alpha=1}^{N}U_{\alpha\alpha\alpha\alpha}.
\end{eqnarray}
Here, $N$ is the number of correlated states that span the target subspace
$\mathcal{C}$. This space is described by the Hamiltonian (\ref{eq:ham}) and
consists typically of d-states localised on specific atoms in the
unit cell.

In cRPA the effective kernel $\mathbf{U}$ is obtained formally from the
effective polarisability
\begin{equation}
\label{eq:splitting}
\chi^r(\omega)=\chi(\omega) - \chi^c(\omega)
\end{equation}
and the bare Coulomb interaction $\mathbf{V}$ via\cite{PhysRevB.70.195104}
\begin{equation}
\label{eq:screening}
\mathbf{U}(\omega)=\mathbf{V}+\mathbf{V}\chi^r(\omega)\mathbf{U}(\omega).
\end{equation}
The effective polarisability $\chi^r$ contains all polarization effects at the
RPA level (described by $\chi$),
except those within the correlated space $\mathcal{C}$. These contributions
are described by the correlated part $\chi^c$.\cite{PhysRevB.70.195104}

We emphasise that this separation is trivial in the Bloch domain only if the
target space $\mathcal{C}$ forms an isolated set of bands known as composite
band.\cite{PhysRevLett.98.046402} In this case, it is always possible to find a
minimal Wannier basis.\cite{PhysRevLett.98.046402} For these systems, the
effective time-ordered polarisability takes the constrained form of the Adler
and Wiser expression\cite{PhysRev.126.413,PhysRev.129.62}
\begin{equation}
\label{eq:unique}
\begin{split}
\chi^c_{\GG\GG'}(\qq,\omega)=&\frac{1}{N_{\kk}}\sum\limits_{\kk}
	\sum\limits_{n,n'\in\mathcal{C}}\frac{(1-f_{n\kk})f_{n'\kk+\qq}}{
	\epsilon_{n\kk}-\epsilon_{n'\kk+\qq}-\omega \mp \ii \eta}\\
&\bra{\phi_{n\kk}}\ee^{\ii(\GG+\qq)\rr}\ket{\phi_{n'\kk+\qq}}\\
&\bra{\phi_{n'\kk+\qq}}\ee^{-\ii(\GG'+\qq)\rr'}\ket{\phi_{n\kk}},
\end{split}
\end{equation}
where $\GG,\GG'$ are reciprocal lattice vectors, $N_{\kk}$ the number of
k-points in the first Brillouin zone,  $f_{n\kk}$ the occupancies of the Bloch
state $\phi_{n\kk}$ having the energy $\epsilon_{n\kk}$ and the shorthand $\mp
\ii \eta = -\ii\eta\ {\rm sgn}(\epsilon_{n\kk}- \epsilon_{n'\kk-\qq})$ was used.
In this simple case, $\chi^r$ includes all transitions but those where the
occupied and filled states are both in the isolated set of bands.  Since the
isolated set of bands is well defined, the procedure is well defined. 

However, the correlated states are often found to be entangled with other
non-correlated ones. As a result, the subspace $\mathcal{C}$, described by the
model Hamiltonian (\ref{eq:ham}), can only be defined in the localised Wannier
basis in most cases. The corresponding expression for the effective
polarisability $\chi^c$ is thus also well defined only in the Wannier basis
\begin{equation}
\label{eq:hanke}
\begin{split}
&\chi^c(\rr,\rr',\omega)=
\frac{1}{N^2_{\kk}}\sum\limits_{nn'\kk\kk'}
\sum\limits_{\balpha\bbeta\bgamma\bdelta\in\mathcal{C}}
	\frac{ (1-f_{n\kk} )f_{n'\kk'}}{
 \epsilon_{n\kk}-\epsilon_{n'\kk+\qq}-\omega \mp \ii \eta}\\
&
T_{\balpha n}^{*(\kk)}
T_{\bbeta n'}^{(\kk')}
w^*_{\balpha}(\rr)
w_{\bbeta}(\rr)
w^*_{\bgamma}(\rr')
w_{\bdelta}(\rr')
T_{\bgamma n'}^{*(\kk')}
T_{\bdelta n}^{(\kk)}
\end{split}
\end{equation}
and follows from Kubo's expression\cite{KaltakThesis}
\begin{equation}
\label{eq:Kubo}
	\chi^c(\rr,\rr',t)=-\ii\left\langle \mathcal{T}\left[\delta n_c(\rr,t)\delta
n_c(\rr',0)\right]\right\rangle_0,
\end{equation}
where $\mathcal{T}$ indicates the Dyson time-ordering operator and $\delta n_c$
the correlated fluctuation density operator measured with respect to
the non-interacting ground state.\cite{StarkeThesis, fetter2003quantum}

In practice, neither Eq. (\ref{eq:Kubo}) nor Eq. (\ref{eq:hanke}) proves to be
practical. The former is overly formal and abstract, whereas the latter is
computationally intractable. Consequently, it is preferable to reformulate
$\chi^c$ in reciprocal space.  A convenient way to find this expression is
presented in the appendix \ref{app:Hanke}. Here we report only the final result
\begin{equation}
\label{eq:chic}
\begin{split}
\chi_{\GG\GG'}^c(\qq,\omega)=&\frac{1}{N_{\kk}}\sum\limits_{nn'\kk}
	\frac{(1-f_{n\kk})f_{n'\kk+\qq}}{
 \epsilon_{n\kk}-\epsilon_{n'\kk+\qq}-\omega \mp \ii \eta}\\
&\bra{\bar\phi_{n\kk}}\ee^{\ii(\qq+\GG)\rr}\ket{\bar\phi_{n'\kk+\qq}}\\
&\bra{\bar\phi_{n'\kk+\qq}}\ee^{-\ii(\qq+\GG')\rr'}\ket{\bar\phi_{n\kk}}
\end{split}
\end{equation}
which contains the correlated Bloch functions and projectors
\begin{align}
\label{eq:barphi}
	\ket{\bar\phi_{n\kk}}=&\sum\limits_{m}P_{nm}^{(\kk)}\ket{\phi_{m\kk}}\\
\label{eq:projector}
	P_{nm}^{(\kk)}=&\sum\limits_{\alpha\in\mathcal{C}}
T_{n \alpha}^{*(\kk)}
T_{\alpha m}^{(\kk)}.
\end{align}

Unfortunately, there is a problem with the definiteness of the polarisability 
$\chi^r$ using the procedures outlined above. This is mostly clearly seen by
inspecting the long-wave length limit of the polarisation, i.e. the value at
$\mathbf{q}\to0$.
To evaluate this, the standard approach is k-p perturbation
theory, which yields for the long-wave limit the following
contribution\cite{AmbroschDraxl20061}
\begin{equation}\label{eq:longwavelimit}
\begin{split}
	\lim_{\mathbf{q}\to0}\frac{\chi(\mathbf{q},\omega) }{|\mathbf{q}|^2} =&
	\frac{1}{N_{\mathbf{k}}}\sum_{nm\mathbf{k}}
	\frac{(1-f_{n\mathbf{k}})f_{
		m\mathbf{k}}}{\epsilon_{n\mathbf{k}} -
	\epsilon_{m\mathbf{k}} - \omega \mp \mathrm{i}\eta}
	|\mathbf{O}_{nm}^{(\kk)}|^2.
\end{split}
\end{equation}
If $\mathbf{H}$ denotes the KS Hamiltonian and $\mathbf{S}$ the
overlap operator the
optical transition elements can be written as\cite{PhysRevB.73.045112}
\begin{equation}\label{eq:OpticalTransitions}\begin{split}
	\mathbf{O}_{nm}^{(\kk)} = &\left\langle\phi_{n\mathbf{k}} \middle|  \nabla_{\mathbf{k}
    }\phi_{m\mathbf{k}} \right\rangle  \\
    =&
	\frac{
\langle\phi_{n\mathbf{k}} \mid \nabla_{\mathbf{k}}
( \mathbf{H} - \mathbf{S} \epsilon_{m\kk})
\mid \phi_{m\mathbf{k}}\rangle
	}{
\epsilon_{n\mathbf{k}} -
	\epsilon_{m\mathbf{k}}
	}
    \end{split}
\end{equation}

In practice, terms with transition energies below a certain
threshold in Eq. (\ref{eq:longwavelimit}) as well as in Eq.
(\ref{eq:OpticalTransitions}) are neglected. This is a robust approach for
the fully screened interaction, but can be problematic when extending the
long-wave limit to cRPA. To describe the $q\to0$ limit in cRPA requires the
subtraction of 
\begin{equation}\label{eq:CorrelatedOpticalTransitions}
	\bar{\mathbf{O}}_{nm}^{(\kk)} =
	\sum_{n',m'} P_{n,n'}^{*(\mathbf{k})}
	O_{n'm'}^{(\kk)}
	P_{m'm}^{(\mathbf{k})}
\end{equation}
from the fully-screened term (\ref{eq:longwavelimit}).
\newline
Unfortunately, the Wannier projection’s disentanglement procedure modifies the
$\mathbf{k}$-dependent wavefunction behaviour at band crossings (prohibited
crossings), such that often more screening terms are subtracted from Eq.
(\ref{eq:longwavelimit}). That is, the positivity condition
\begin{equation}
\label{eq:positivity}
|\bar{\mathbf{O}}_{nm}^{(\kk)}|^2 \le |\mathbf{O}_{nm}^{(\kk)}|^2
\end{equation}
can be violated for bands in the correlated subspace that are subject to Wannier
projection. The result are negative definite dielectric functions, and with that
negative cRPA interactions at zero frequency.
\newline

This undermines the method’s reliability and is the main reason why several
approximations were proposed in the community that we summarise in the
following. Finally, we also present a new scheme: s-cRPA that remedies most of
the shortcomings of the devised schemes.

\subsubsection{Weighted method: w-cRPA}\label{sec:weighted}
The scheme of \sais et al., known as w-cRPA,\cite{Merkel2024} follows from Eq.
(\ref{eq:chic}) using\cite{PhysRevB.83.121101, PhysRevB.85.045132}
\begin{equation}
\label{eq:weight}
	P_{nm}^{(\kk)} \to \sqrt{	P_{nn}^{({\kk})}} \delta_{nm}.
\end{equation}
The diagonal of the projectors measures the {\em correlation degree}
of all Bloch bands that contribute to the Wannier states at a specific k-point
and can be understood in terms of the statistical leverage score for band $n$.

The statistical leverage measures how {\em outlying} a row is with respect to
the span $N$ of the matrix.\cite{james2013} More precisely, the leverage score
of row $n$ of $P$ is defined via
\begin{equation}\label{eq:leverage}
	l_n = \|v_{n}\|^2_{N}, \quad 0\le l_n \le 1,
\end{equation}
were $v_n$ is the $n$-th singular eigenvector of $P$ and the norm is computed
over the first $N$ dimensions of $v_n$, corresponding to the span of
$P$.\cite{Drineas2012} These vectors constitute the columns of the unitary
matrix $V$ in the singular value decomposition $P=U\Sigma V^\dagger$.  In the
appendix \ref{app:Leverage} we show that $l_n=P_{nn}$ holds true at each
k-point. 

The w-cRPA method has two favourable properties. First, the positivity condition
(\ref{eq:positivity}) is never violated; the leverage is a positive number
between 0 and 1.  Second, w-cRPA conserves the number of electrons, because the
trace of the projectors remains unchanged by the approximation
(\ref{eq:weight}).

Nevertheless, the robustness of the method comes with a caveat.  Screening
effects inside the correlated subspace described by the off-diagonal terms of
the correlated projectors (\ref{eq:projector}) are not removed from the fully
screened polarisability in w-cRPA.  The interaction in resulting model
Hamiltonians is often too weak to describe important phase
transitions.\cite{Merkel2024,PhysRevB.111.195144} A cRPA method that takes
off-diagonal components of the projectors (\ref{eq:projector}) into account and
thus removes more screening effects from the correlated subspace is discussed in
the following.

\subsubsection{Projector method: p-cRPA}\label{sec:projector}
From the discussion at the end of section \ref{sec:crpa} one concludes that
preserving the positivity condition (\ref{eq:positivity}) without neglecting
screening effects, that is, off-diagonal components in the correlated projectors
(\ref{eq:projector}), are conflicting goals.  To solve this dilemma at least
partially, one can sacrifice electron conservation and delete almost linearly
dependent entries in $P$. It can be expected that almost linearly-dependent
correlated Bloch states are close in energy and thus the main reason for the
violation of (\ref{eq:positivity}).  In p-cRPA, these vectors are identified as
the zeros of the eigenvalues $\Theta_n$ of the projector after a Jacobi
diagonalisation\cite{Cardoso_Souloumiac, NR2007}
\begin{equation}\label{eq:p-crpa}
P_{nm}^{(\kk)} \to  \Theta_{n}^{(\kk)}P_{nm}^{(\kk)}.
\end{equation}
Note, the original projector $P$ is idempotent and thus has a discrete spectrum
$\Theta_n^{(\kk)} \in \lbrace 0,1\rbrace$. This is no longer the case for p-cRPA, since
$P^2\neq P$, which is equivalent to sacrificing electron conservation. 
It has been shown that the regularisation (\ref{eq:p-crpa})
improves the robustness of Eq. (\ref{eq:chic}) significantly when combined with
the long-wave limit (\ref{eq:longwavelimit}) and
(\ref{eq:CorrelatedOpticalTransitions}).\cite{KaltakThesis}

The eigenvalue ordering of the Jacobi diagonalisation algorithm closely
follows the value of the leverage; column vectors with the smallest leverage are
removed. However, the ordering is not always exact, especially for dense k-point
meshes the algorithm sometimes fails to select the Bloch bands with highest
leverage score.  Imposing strict ordering and selecting columns with highest
leverage score gives rise to a revised version of p-cRPA, called p-cRPA-rev in
the following.

Both methods, p-cRPA and p-cRPA-rev remove more screening effects compared to
w-cRPA, because off-diagonal terms in the projector are present, even after
regularisation. Unfortunately, there are cases where this regularisation still
introduces negative long-wave limits as demonstrated in Fig. \ref{fig:kdep}
of section \ref{sec:results}. In such cases, positive definite interactions are
only obtained for very dense k-point samplings. We suspect that the origin of
this problem stems from the fact that electron conservation is violated, i.e.
the trace of the "projector" is not conserved after regularisation.

\subsubsection{Spectral method: s-cRPA}\label{sec:spectral}
To remedy the drawbacks of p-cRPA and w-cRPA we propose to use the spectrum
of the original projectors (\ref{eq:projector})
\begin{equation}
\label{eq:s-crpa}
P_{nm}^{(\kk)} \to \Theta_{n}^{(\kk)}\delta_{nm},
\end{equation}
ordered by the leverage (\ref{eq:leverage}). We denote this method in the
following as spectral cRPA or s-cRPA due to its connection to the eigenspectrum
of the original projector.

This cRPA scheme is a mixture of w-cRPA and p-cRPA(-rev) with altered weights
(\ref{eq:weight}) that includes also information about off-diagonal components
of the original projectors.  Since the trace of Eq. (\ref{eq:s-crpa}) is
conserved (as for w-cRPA), the corresponding long-wave limit is strictly
positive definite; a fact that we also observed for all our calculations
reported in the section (\ref{sec:results}).

The scheme suggested here is conceptually exceedingly simple. Eq.
(\ref{eq:s-crpa}) either includes a specific Bloch state in the calculation of
the polarisability of the correlated subspace, or it entirely neglects a state.
At each k-point, the same number of $N$ {\em target} states is selected
(typically $N=5$ states for d-electrons). The key differentiator of the present
method is that the choice of whether a state is included or not is solely based
on the leverage score; only those Bloch states are selected that involve the
strongest contribution from Wannier orbitals, and Bloch states that involve the
weakest contribution of Wannier states are neglected. This is in the spirit of
the CUR rank compression.\cite{Mahoney2009} That is, one does not seek a linear
combination of states (as done for instance, in singular value decompositions),
but rather selects at each k-point the most relevant active Bloch states. It
should be clear that this simply removes contributions from the full
polarisability that were initially included. Hence, positive definiteness of the
residual polarisability is guaranteed. Furthermore, the number of states is
exactly conserved. Despite its simplicity, we will show below that the results
of this method follow all expected trends, and the method also leads to fast
k-point convergence.

\subsection{Computational details}\label{sec:details}
We have examined the effective interaction for Scandium (Sc) and Copper (Cu) in
face-centred cubic (fcc) lattices with lattice constants of 4.64~\AA~ and
3.52~\AA, respectively. Given the metallic nature of these systems, Brillouin
zone integration can be challenging, particularly when flat and partially
occupied bands are present. To address this issue, we employed a methodological
approach that involved determining the ground state wavefunction and the
long-wave limit (\ref{eq:longwavelimit})
using a dense k-point grid of $24\times24\times24$ points with a smearing factor
of $\sigma=0.1$ eV$^{-1}$. The same k-point grid was utilised for the Wannier
projection, while a coarser grid of $8\times8\times8$ points was employed during
the cRPA step. This k-point grid was also used to determine the spatial decay of
$U$ up to 14 \AA.

To achieve convergence of the effective interaction matrix with respect to the
plane wave basis set, we followed the method described in
Ref.~\onlinecite{PhysRevB.86.085117}. This approach involves determining the
effective matrix elements using a low energy cutoff $E^{\rm low}_{\rm cut}$ for
the cRPA polarisability $\chi^r$ on a plane wave grid using shape restoration of
the density.\cite{LMAXFOCKAE} The result is corrected by the difference between
the bare Coulomb interaction matrix obtained with the same technique and the one
obtained using exact one-centre terms for the charge density with a higher
cutoff $E_{\rm cut}^{\rm high}$. This correction is justified because
high-energy contributions to the polarisability $\chi^r$ vanish, and screening
becomes ineffective for large $\GG$ vectors.

Specifically, we used $E_{\rm cut}^{\rm low}=400$ eV for the cRPA
polarizability, $E_{\rm cut}^{\rm high}=700$ and $E_{\rm cut}=500$ eV for the
PAW basis set.  Similar to $GW$ calculations, all bands provided by the plane
wave cutoff are included in cRPA calculations.  For our chosen settings,
this corresponded to 288 bands for the transition metals.

We also explored the use of a mixed stochastic-deterministic compression
algorithm to reduce the number of bands.\cite{Altman2024} We employed a constant
energy ratio of $F=0.04$, with $N_p=32$ protected bands and two stochastic bands
per pseudo state $(N_\xi=2)$, resulting in a total of 106 bands. This approach
achieved a speed-up of approximately 50 per cent for the cRPA calculations while
incurring a negligible error of less than 50 meV in the effective interaction
matrix. However, we found that naive truncation of the basis set without any
compression yielded a similar level of error. Consequently, we recommend
avoiding this compression for most cRPA calculations, except perhaps for very
challenging (almost norm-conserving) pseudopotentials.

All results have been obtained within the finite-temperature formalism of
many-body perturbation theory on the imaginary frequency axis. For this purpose,
a compressed Matsubara grid has been determined by preserving the isometry of
the time and frequency representations of Green's functions in
\texttt{NOMEGA}=24 dimensional subspaces.\cite{Kaltak2020} Typically, half as
many points suffice to determine the static effective interaction. However,
doubling the number of points is helpful to reconstruct details of the
real-frequency interactions using a Pad\'e fit of the data on the imaginary
axis.  For this purpose, we have applied the adaptive Antoulas-Anderson (AAA)
algorithm\cite{AAA2018} as interfaced in \texttt{py4vasp}.\cite{py4vasp}

\section{Results}\label{sec:results}
\subsection{Wannier basis for Sc and Cu}
\begin{figure}
\includegraphics[width=0.45\textwidth]{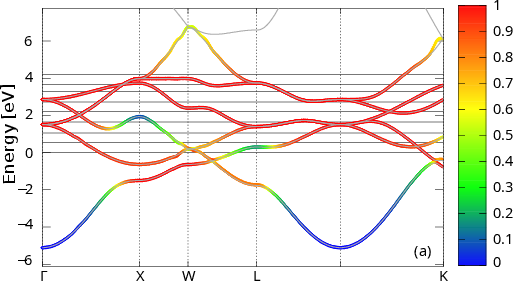}
\includegraphics[width=0.45\textwidth]{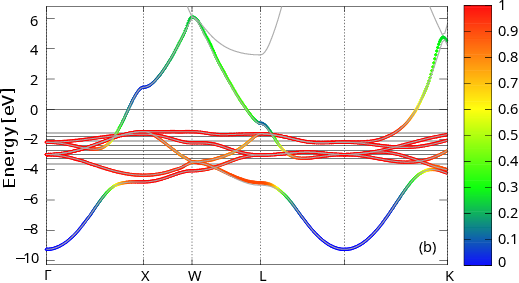}
	\caption{Bands of fcc Sc (a) and Cu (b) with d-character resolution obtained
	from \texttt{Wannier90}.\cite{Mostofi2008685} Red bands indicate strong
	d-character, blue bands indicate strong s-character. For reference, the
	original band structure obtained with \texttt{VASP}\cite{PhysRevB.59.1758} is
	shown in grey.  Filling levels are indicated by horizontal lines (neutral
	compound at zero energy).
}
\label{fig:bands}
\end{figure}
The effectiveness of the cRPA calculations is significantly influenced by the
quality of the Wannier basis. For 3d transition metals, the absence of an
isolated set of bands precludes a bijective mapping between Bloch bands and
Wannier functions. Nevertheless, an approximate one-to-one correspondence can be
established between six Wannier states and six bands in the vicinity of the
Fermi level. These bands comprise five narrow bands, predominantly of
d-character, and one itinerant band capturing the 4s orbital.

In our study, we positioned the s-like Wannier state at the direct coordinates
$(0.25, 0.25, 0.25)$, while the d-states were centred on the transition metal
atom at the origin. Figure \ref{fig:bands} illustrates a comparison between the
original Kohn-Sham bands and the eigenvalues of the Hamiltonian in the Wannier
basis for both Sc and Cu. The Wannier basis demonstrates high accuracy in
reproducing most symmetry lines; however, it encounters difficulties in
representing the high symmetry points W and K of the s-state.  This limitation
arises from these points being intersections with higher energy states,  which
are not considered in our Wannier basis. Consequently, the eigenvalues of the
Wannier Hamiltonian exhibit oscillatory behaviour at these points.

Given that the representability of the Wannier basis is inherently imperfect, we
implemented a very dense k-point grid of $24\times 24\times 24$ for the Wannier
projection to allow the aforementioned disentanglement of states in the Wannier
construction to work optimally.

\subsection{Electron filling and Hubbard $U$}
To evaluate the impact of the cRPA methods at various electron fillings, we
froze the Wannier projection while adjusting the electron number—increasing it
for Sc and decreasing it for Cu. In Table \ref{tab:U-doped-Sc} and Table
\ref{tab:U-doped-Cu} we present the effective averaged on-site interaction $U$
(see Eq.~\ref{eq:Udef}) in eV for Sc and Cu at different fillings.
\begin{table}
\begin{center}
  \caption{Effective averaged on-site interaction $U$ (see Eq.~\ref{eq:Udef}) in
	eV for Sc at different fillings, comparing the different disentanglement schemes
	presented in Sec.~\ref{sec:crpa}.}
\begin{tabular*}{1.00\columnwidth}{@{\extracolsep{\fill}}lccccccccc}
\hline
\hline
cRPA & Sc         & Sc$^-$     & Sc$^{2-}$ & Sc$^{3-}$ & Sc$^{4-}$ & Sc$^{5-}$ & Sc$^{6-}$ & Sc$^{7-}$ & Sc$^{8-}$ \\
\hline
w     & 2.4  & 2.3 & 2.3  & 2.6  & 2.6  &  2.6 &  2.4 &  2.1 & 1.9 \\
p     & 2.2  & 1.9 & 2.0  & 2.4  & 2.7  &  2.8 &  2.6 &  2.5 & 2.2 \\
p-rev & 2.2  & 1.9 & 2.0  & 2.4  & 2.7  &  2.8 &  2.6 &  2.4 & 2.1 \\
s     & 2.6  & 2.4 & 2.6  & 2.8  & 3.0  &  3.0 &  2.8 &  2.6 & 2.4 \\
\hline
\hline
\end{tabular*}
\label{tab:U-doped-Sc}
\end{center}
\end{table}

\begin{figure}
\includegraphics[width=0.45\textwidth]{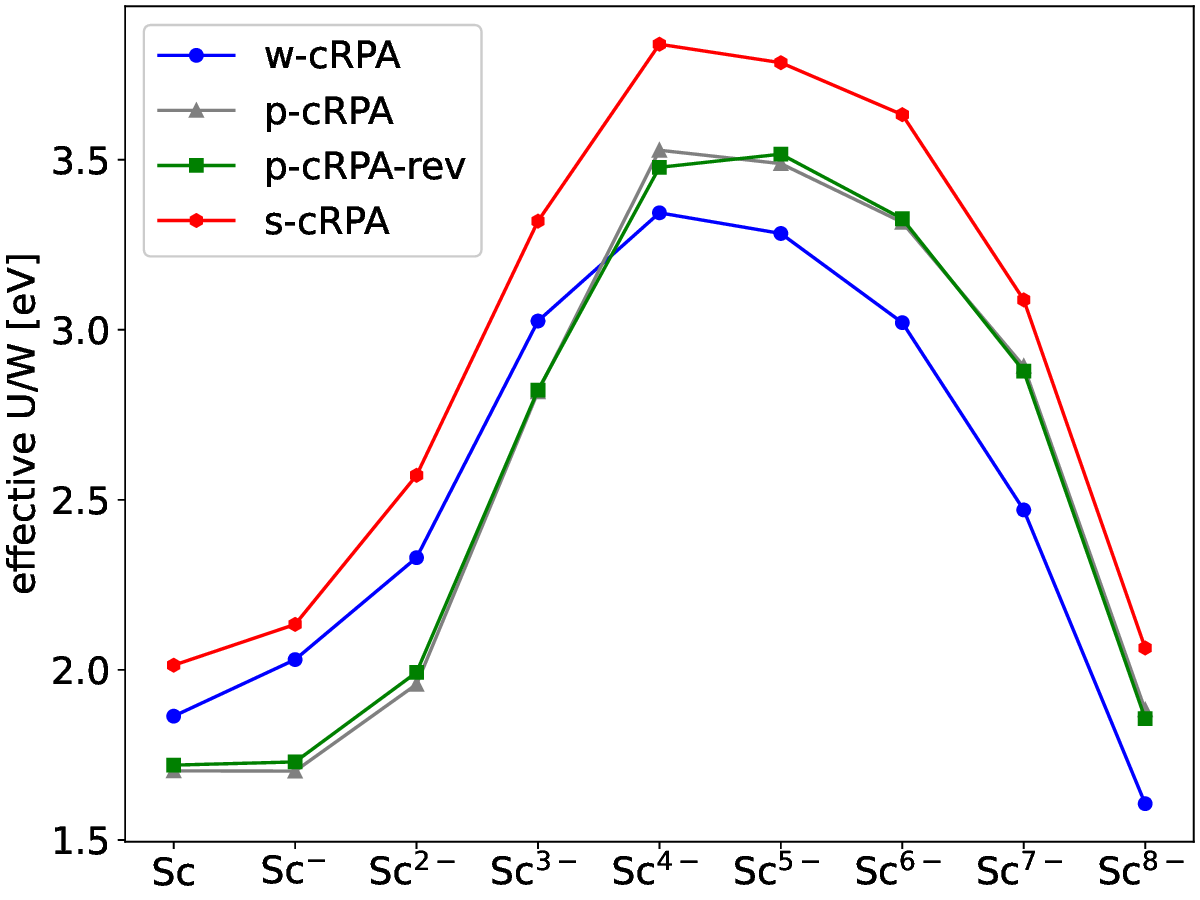}
\includegraphics[width=0.45\textwidth]{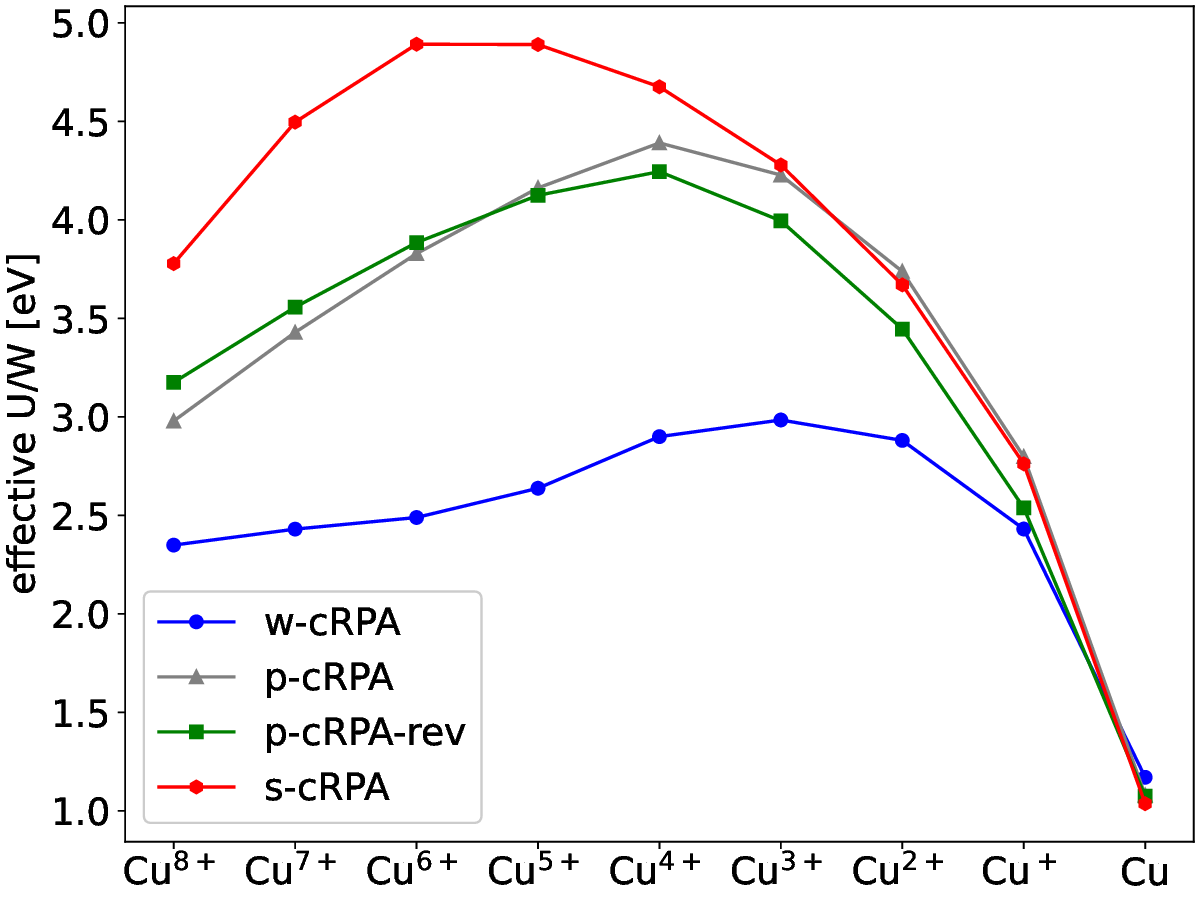}
\caption{Comparison of the quotient of on-site effective d-Coulomb repulsion $U$
	and fully screened interaction $W$ for Sc and Cu at different fillings with the
	different disentanglement schemes presented in Sec.~\ref{sec:crpa}.  }
\label{fig:cRPA}
\end{figure}

Additionally, we analysed the quotient $U/W$, where $W$ represents the fully
screened Hubbard-Kanamori interaction. This approach allows us to isolate and
compare the unscreening effect of each method without incorporating Wannier
localisation effects. The results of this comparison are presented in Figure
\ref{fig:cRPA}.
\begin{table}
	\centering
\caption{Effective averaged on-site interaction $U$ (see Eq.~\ref{eq:Udef}) in
	eV for Cu at different fillings, comparing the different disentanglement
	schemes presented in Sec.~\ref{sec:crpa}. ($^\diamond$) values obtained from
	extrapolation without long-wave limit, see Fig. \ref{fig:kdep}.}
	\begin{tabular*}{1.00\columnwidth}{@{\extracolsep{\fill}}lccccccccc}
	\hline
	\hline
	cRPA & Cu$^{+8}$  & Cu$^{+7}$  & Cu$^{+6}$ & Cu$^{+5}$ & Cu$^{+4}$ & Cu$^{+3}$ & Cu$^{+2}$ & Cu$^{+}$  & Cu \\
	\hline
	w     & 3.3 & 3.0 & 2.8 & 2.9 & 3.2 & 3.4 & 3.5 & 3.6 & 4.7 \\
	p     & 4.2 & 4.2 & 4.3 & 4.6 & 4.9 & 4.8 & 4.6 & 4.2 & 4.3$^\diamond$ \\
	p-rev & 4.5 & 4.4 & 4.4 & 4.6 & 4.7 & 4.5 & 4.2 & 3.8 & 4.3$^\diamond$ \\
	s     & 5.3 & 5.5 & 5.5 & 5.4 & 5.2 & 4.9 & 4.5 & 4.1 & 4.1 \\
	\hline
	\hline
	\end{tabular*}
	\label{tab:U-doped-Cu}
\end{table}

For all calculations the s-cRPA gives the largest interactions across all
fillings both for Sc and Cu. The p-cRPA and p-cRPA-rev methods produce
intermediate results, often very close to each other. This similarity
demonstrates that the employed Jacobi diagonalisation method effectively orders
the eigenvalues according to the $N$-dimensional leverage score.

All methods exhibit their maximum effect roughly at half-filling, which is
expected due to the prevalence of intra d-d transitions that are removed from
screening.  The maximum is more pronounced for Sc compared to Cu, likely due to
the greater localisation of d-states in Cu. The latter is also the reason for
the overall larger interaction values compared to the Sc series.

Figure \ref{fig:cRPA} illustrates that s-cRPA removes most intra-d screening
effects, more so than p-cRPA and p-cRPA-rev. This difference is attributed to
the electron number conservation in s-cRPA, which is not maintained in the
regularisation scheme employed in p-cRPA. Both p-cRPA-(rev) and s-cRPA
implicitly consider off-diagonal elements of the correlated projector,
whereas w-cRPA neglects these contributions, resulting in the smallest
interactions overall.

\subsection{Frequency dependence}
\begin{figure}
\includegraphics[width=0.45\textwidth]{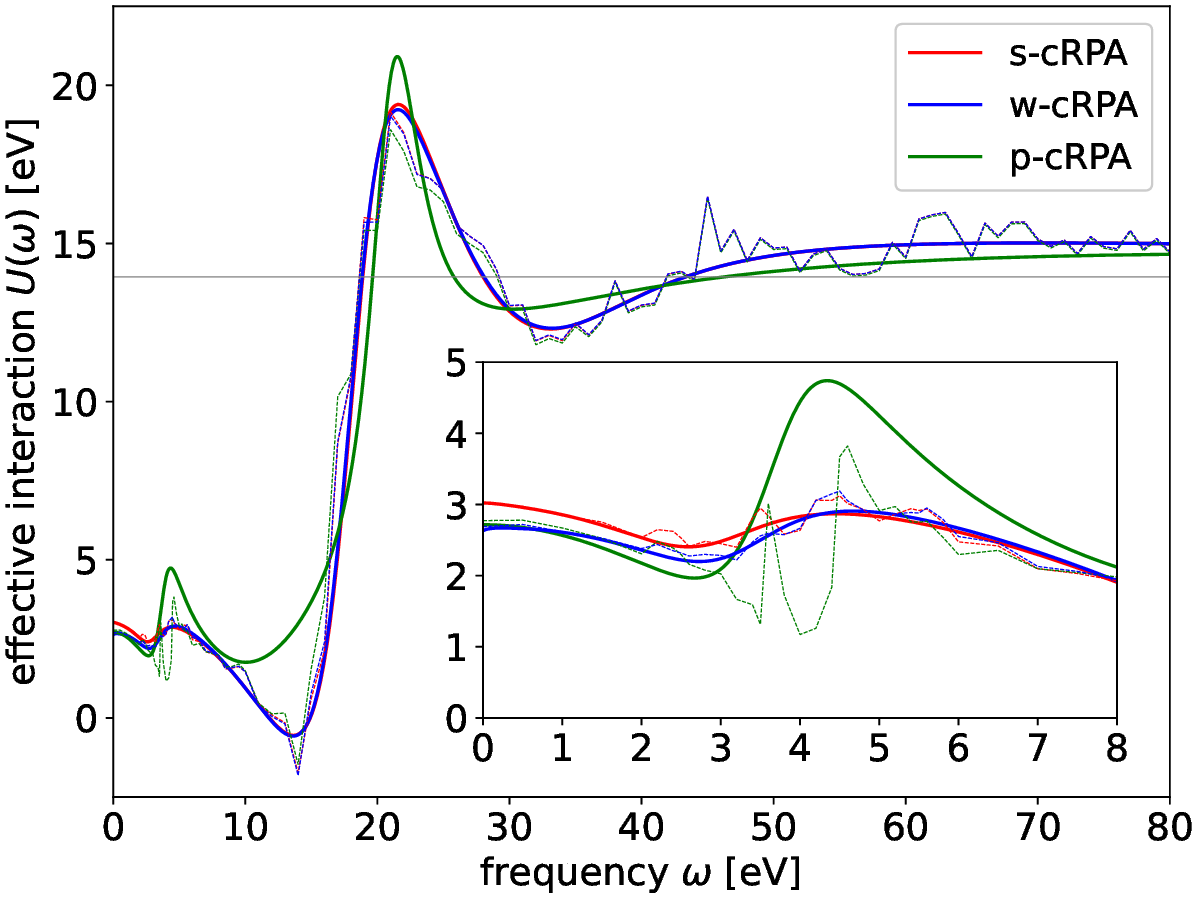}
\caption{(solid) Frequency dependence of effective interaction $U$ for
	Sc$^{4+}$ determined with various cRPA methods obtained from analytic
	continuation using a compressed Matsubara grid of 24 points.\cite{Kaltak2020}
	For comparison data obtained from  direct computation on the real-frequency
		axis is shown (dashed) Gray line corresponds to bare Coulomb interaction.
		}
\label{fig:Uomega}
\end{figure}
Next, we focus on the frequency dependence of the effective one-site interaction
for Sc$^{4+}$. Figure \ref{fig:Uomega} shows $U$ as a function of the frequency
$\omega$ for s-cRPA, w-cRPA and p-cRPA obtained from analytic continuation of
the compressed Matsubara data to the real axis (solid lines).\cite{Kaltak2020}
For comparison real-frequency data obtained from direct, point-wise calculation
is also shown (dashed lines). As can be seen, reconstruction of real-frequency
data is excellent, including the plasmon peak around 20 eV. Even unique
features, such as the peak of p-cRPA around 4.5 eV are resolved well by the AAA
fit.\cite{AAA2018}

All methods approach the bare Coulomb interaction (grey horizontal line) at high
frequencies around 150 eV (not shown), but differ substantially at low
frequencies. The inset shows the pronounced peak for p-cRPA at around 4.5 eV
(green line) that is missing for w-cRPA and s-cRPA, but present for the fully
screened interaction (not shown). Overall, s-cRPA and w-cRPA frequency
dependence is similar, being almost constant at low frequencies and deviating
substantially from the low-frequency regime of p-cRPA.

\subsection{Spatial decay of Hubbard $U$}
\begin{figure}
\includegraphics[width=0.45\textwidth]{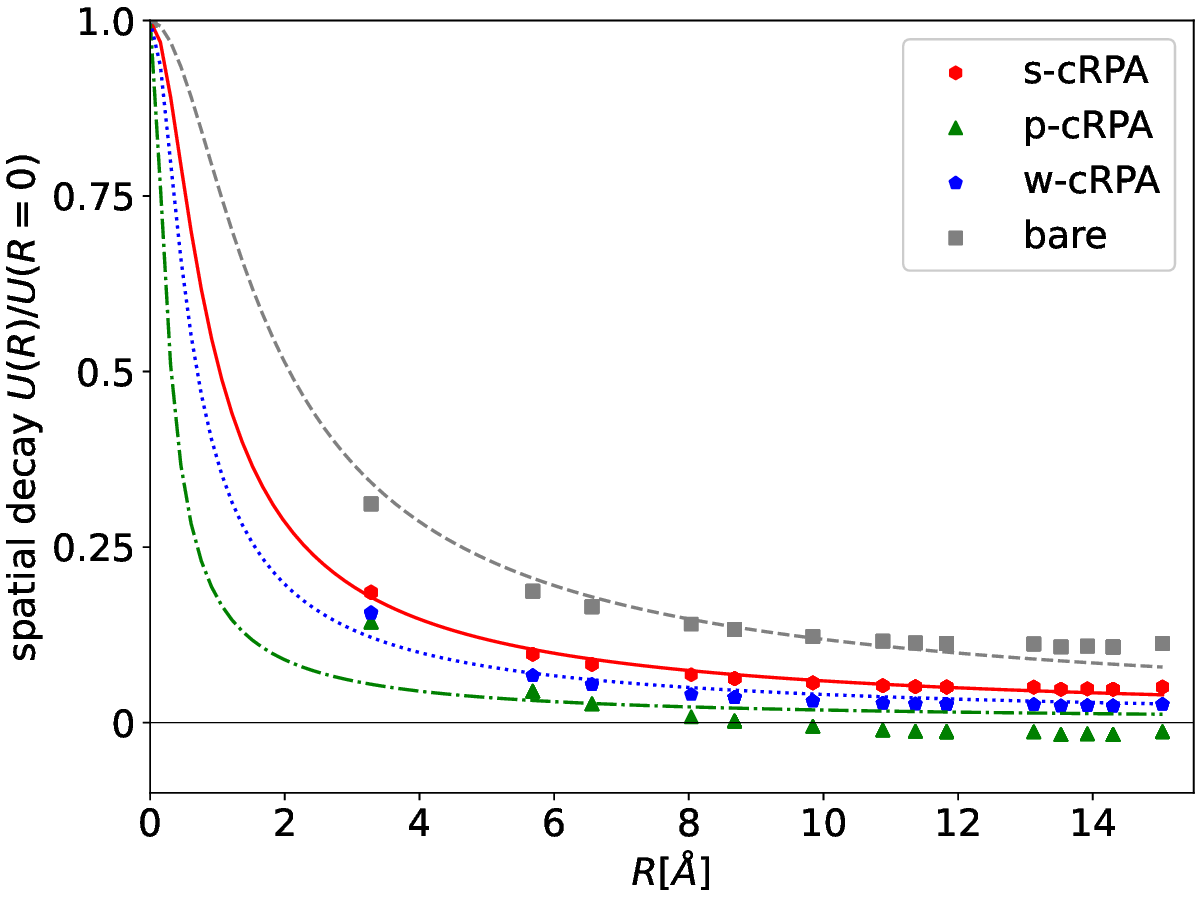}
\caption{Spatial decay of effective cRPA interactions scaled to one-site
	interaction $U(R)/U(R=0)$ for Sc$^{4-}$.
}
\label{fig:Rdep}
\end{figure}
We shift the focus to the spatial decay of the effective cRPA interactions that
is often described by the Ohno potential\cite{Reddy2024,DiSante2023}
\begin{equation}
	\frac{U(R)}{U(R=0)} = \frac{1}{\sqrt{\frac{R}{\delta}+1}}.
\end{equation}
To this end, we have fitted this model to the calculated cRPA interactions
\begin{equation}
	U_{\bar\alpha\bar\beta\bar\gamma\bar\delta}=\bra{\pw_{\bar\alpha},
	\pw_{\bar\beta}}\mathbf{U}(\omega=0)\ket{\pw_{\bar\delta},\pw_{\bar\gamma}},
\end{equation}
restricted to two centres $\RR_\alpha=\RR_\delta=\RR$ and
$\RR_\beta=\RR_\gamma=0$. The results are shown in Fig. \ref{fig:Rdep}.
Overall, it can be seen that the cRPA interactions decay faster than the bare
Coulomb interaction (squares). The Ohno model describes the decay of the s-cRPA
interaction almost perfectly (hexagons), with slight deviations for w-cRPA for
the nearest neighbour site (pentagons). Both, s-cRPA and w-cRPA approach each
other with increasing distance.  This is in contrast to p-cRPA. Judging from
the figure, it seems that p-cRPA decays faster than the Ohno potential
(triangles).  However, trying to fit this behaviour with another model is most
probably misleading, since the interaction becomes negative around 8 \AA.  It
can be expected that this nonphysical behaviour stems from the fact that this
method violates particle number conservation and thus should be avoided in any
cluster expansions of the model Hamiltonian.

\subsection{$\kk$-point convergence}
\begin{figure}
\includegraphics[width=0.45\textwidth]{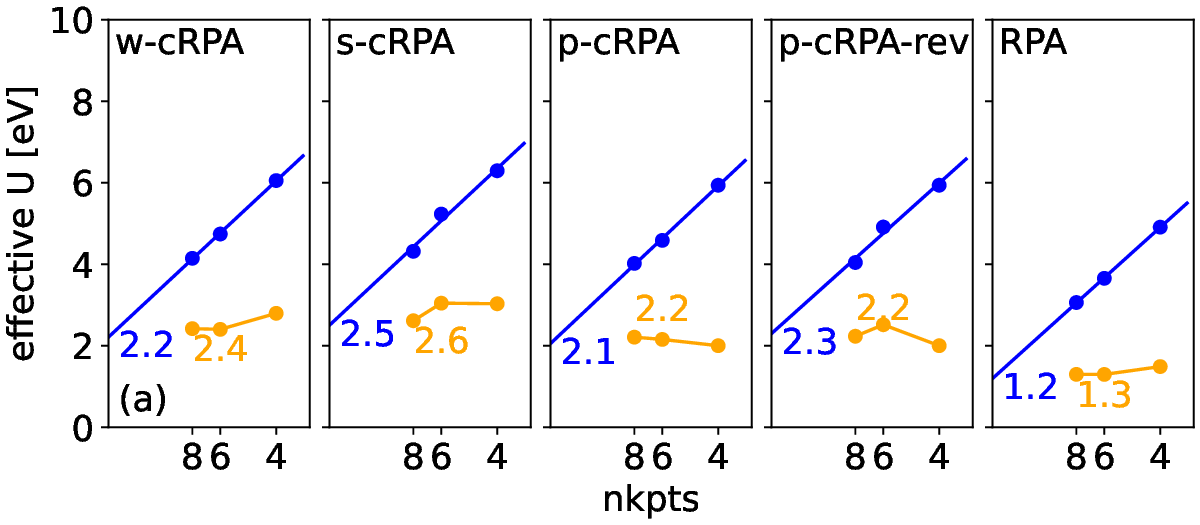}
\includegraphics[width=0.45\textwidth]{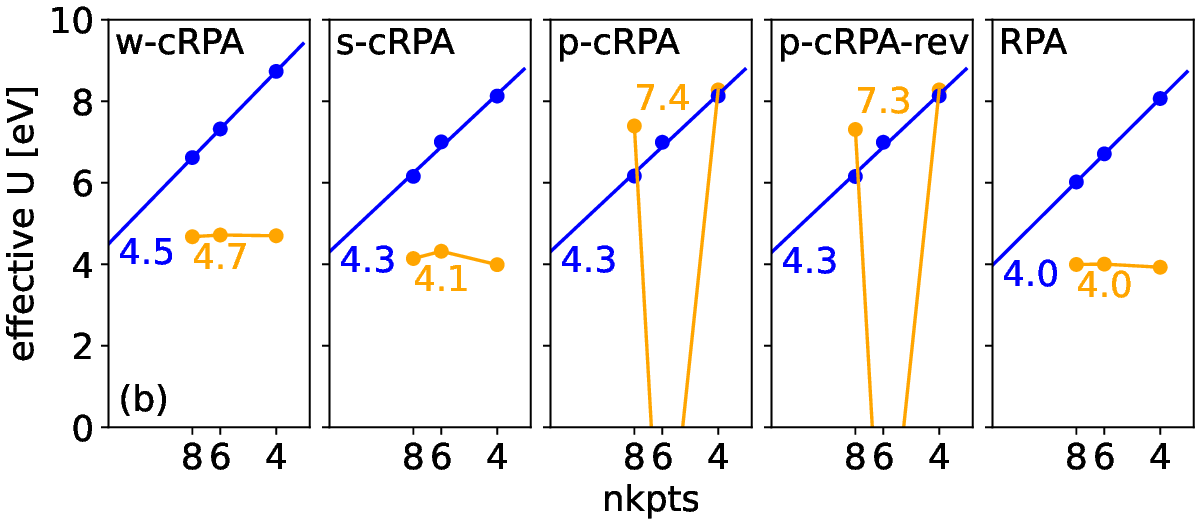}
\caption{
	On-site Hubbard-Kanamori parameter $U$ (\ref{eq:Udef}) as a function of
	k-point sampling for each cRPA method studied, along with the fully screened
	interaction in RPA for Sc (a) and Cu (b).
	Each series contains four plots, one for each method. In
	every plot, the x-axis represents the number of k-points per direction used to
	sample the first Brillouin zone, and the y-axis shows the calculated $U$
	value. The blue line corresponds to data obtained without the long-wavelength
	limit correction; it exhibits a linear dependence on the inverse number of
	k-points, allowing for extrapolation to infinite k-point
	sampling.\cite{PhysRevB.111.195144}  The extrapolated interaction value
	is indicated as a blue number within each plot.  The orange line shows results
	including the long-wavelength limit correction, with the value at an
	$8\times8\times8$ k-point grid represented by an orange number.
	}
\label{fig:kdep}
\end{figure}
We analyse the k-point convergence of the effective interaction $U$ by comparing
results obtained without the long-wave limit correction, where $U$ depends
linearly on the inverse number of k-points used, as described in a recent
study.\cite{PhysRevB.111.195144} This linear dependence is illustrated in
Fig.~\ref{fig:kdep} (blue lines) for Sc and Cu across all cRPA methods
considered, including the fully screened interaction, and allows extrapolation
to infinite k-point sampling (blue numbers).  Comparing these with the results
that incorporate the $q \to 0$ limit (shown in orange) reveals that including
this correction (Eq.~\ref{eq:longwavelimit}) substantially improves k-point
convergence in almost all cases. The main exceptions are the projector cRPA
methods (p-cRPA and p-cRPA-rev) applied to Cu at a $6 \times 6 \times 6$ k-point
mesh, where incorporating the long-wave limit leads to negative interaction
values, rendering these approaches unreliable in practice. This is due to the
violation of the positivity condition Eq. (\ref{eq:positivity}). In contrast, the
w-cRPA and s-cRPA demonstrate markedly greater stability with respect to k-point
sampling, comparable to that of the fully screened interaction. These latter two
approaches converge more smoothly with the inclusion of the $q \to 0$ limit,
making them more stable and dependable for practical use. 
\subsection{Effective interaction for CaFeO$_3$}
\begin{table}
	\centering
			\caption{
				Effectively screened, averaged on-site interaction $U$, exchange
	$J$ and nearest neighbour interaction $V$ of CaFeO$_3$ in localised (frontier) basis
	as calculated in Ref. \onlinecite{Merkel2024} and s-cRPA (present work).}
	\begin{tabular*}{1.00\columnwidth}{@{\extracolsep{\fill}}lccc}
	\hline
	\hline
	                 & p-cRPA      & w-cRPA         & s-cRPA       \\
	\hline
		$U$             & 1.25 (1.75) &  0.45 (1.75)   & 2.51 (1.90)  \\
		$J$             & 0.92 (0.53) &  0.80 (0.53)   & 0.97 (0.60)  \\
		$V$         & 0.03 (0.53) &  0.02 (0.53)   & 0.53 (0.53)  \\
	\hline
	\hline
	\end{tabular*}
	\label{tab:CaFeO3}
\end{table}
Finally, we examined the averaged effective interaction for CaFeO$_3$, as
previously explored by Merkel and Ederer.\cite{Merkel2024} In their recent
study, they employed p-cRPA and w-cRPA methods---each within both localised and
frontier basis sets---to determine the interaction parameters $U$ and $J$ for
DFT+$U$ calculations. The precise definitions of these parameters, which differ
from the Kanamori form used in Eq.~(\ref{eq:Udef}), are detailed in their
publication.\cite{Merkel2024} In Table~\ref{tab:CaFeO3}, we present only the
final s-cRPA results, alongside a comparison to the onsite, exchange and nearest
neighbour interaction values reported by Merkel and Ederer for p-cRPA and w-cRPA,
respectively. It is evident from these data that s-cRPA produces the largest
interaction strengths among the considered methods. Nevertheless, even the
onsite value of $2.51\,\mathrm{eV}$ predicted by s-cRPA would be insufficient
for an accurate description of the $P2_1/n$ phase, as it would yield a metallic
ground state with an almost vanishing $R_1^+$ mode. This is at odds with
experimental observations, where an insulating phase stabilises, that is
predicted with DFT+$U$ only within the range $4 \le U \le
7$\,eV.\cite{Merkel2024} 

Achieving a more realistic value for $U$ may require combining s-cRPA with
beyond RPA methods\cite{PhysRevB.104.045134}	or alternative approaches such as
the moment-constrained RPA method proposed by Spencer and
Booth,\cite{PhysRevLett.132.076401} which both have shown promise in refining
effective interaction parameters beyond standard cRPA.

\section{Conclusion}\label{sec:conclusion}

In conclusion, our analysis of various constrained Random Phase Approximation
(cRPA) methods applied to 3d transition metals provides valuable insights into
their relative performance and applicability. The results indicate that the
spectral cRPA (s-cRPA) method offers several advantages for calculating
effective interactions in correlated materials.

The s-cRPA method demonstrates favourable characteristics, including the
conservation of electron number—a physically important constraint that
contributes to the consistency of computed results. 
By incorporating off-diagonal elements of the correlated projector, this
approach effectively eliminates the majority of intra-d screening contributions,
thereby addressing the double counting problem inherent in the treatment of
strongly correlated d-electrons within post mean-field methods.

Significantly, the application of s-cRPA to the realistic compound CaFeO$_3$
yields interaction parameters closer to those required in DFT+$U$ calculations
to correctly reproduce the experimentally observed insulating state. This
highlights the practical advantage of s-cRPA for capturing the electronic
correlations responsible for the gap formation in complex materials.

Recent enhancements to our implementation enable calculation of multi-centre
interactions, allowing detailed study of the spatial decay of the screened
Hubbard interaction U. We have also developed a low-scaling variant that
utilises a compressed Matsubara frequency grid, facilitating efficient
computation of the full frequency dependence of effective interactions in large
systems and enabling studies of defect effects on the effective interaction.

Based on these findings and methodological improvements, s-cRPA appears to be a
promising approach for calculating effective Hubbard interactions in transition
metals and related strongly correlated systems. It combines several desirable
features while addressing some limitations of alternative methods, offering a
balanced approach that merits consideration for applications in condensed matter
physics and materials science. Future work should focus on applying s-cRPA to a
broader range of materials and systematically comparing predictions with
experimental data.
\begin{acknowledgments}
B.K. acknowledges support from the National Research Foundation of Korea (NRF;
	Grants No.  NRF2021R1A4A1031920, No. RS-2021-NR061400, and No. RS2022-NR068223)
	and KISTI Supercomputing Center (Project No. KSC-2023-CRE-0413).
\end{acknowledgments}
\appendix
\section{Derivation of Eq. \ref{eq:chic}}\label{app:Hanke}
A derivation of Eq. (\ref{eq:chic}) starting from the Kubo formalism
(\ref{eq:Kubo}) can be found elsewhere.\cite{KaltakThesis} Here, we derive  Eq.
(\ref{eq:chic}) starting from the Wannier representation Eq.
(\ref{eq:hanke}).\footnote{Eq. (\ref{eq:hanke}) is the constrained version of
full RPA polarisability in Wannier basis originally derived by Hanke and
Sham\cite{PhysRevB.21.4656}.} The expression is obtained by inserting the
inverse Wannier transformation
\begin{equation}
\label{eq:invtransformation}
\ket{\phi_{n\kk}}=\frac{1}{N_{\kk}}\sum\limits_{\alpha\RR}\ee^{-\ii\kk\RR}
	T_{n\alpha }^{*(\kk)}\ket{w_{\alpha\RR}}
\end{equation}
into Eq. (\ref{eq:hanke}) and performing the Fourier transformation to
reciprocal space yielding
\begin{widetext}
\begin{equation}\label{eq:explicit}
\begin{split}
\chi^c_{\GG\GG'}(\qq,\omega)=&
\frac{1}{N_{\kk}^2}
\sum\limits_{\bn \bn'}
\frac{1}{N_{\kk}^4}
\sum\limits_{\bn_1\bn_2\bn_3\bn_4 }
	\frac{(1-f_{\bn})f_{\bn'}}{
	\epsilon_{\bn}-\epsilon_{\bn'}-\omega \mp\ii\eta
	}
\bra{\phi_{\bn_1}}\ee^{\ii(\GG+\qq)\rr}\ket{\phi_{\bn_2}}
\bra{\phi_{\bn_3}}\ee^{-\ii(\GG'+\qq)\rr'}\ket{\phi_{\bn_4}}\\
\times&
\sum\limits_{\alpha\beta\gamma\delta\in\mathcal{C}}
T^{*(\kk)}_{n \alpha}
T^{(\kk_1)}_{\alpha n_1}
T^{*(\kk_2)}_{n_2 \beta}
T^{(\kk')}_{\beta n'}
T^{*(\kk')}_{n'\gamma}
T^{(\kk_3)}_{\gamma n_3}
T^{*(\kk_4)}_{ n_4\delta}
T^{(\kk)}_{\delta n}\\
\times&\sum\limits_{\RR_\alpha}
\ee^{\ii\RR_\alpha(\kk_1-\kk)}
\sum\limits_{\RR_\beta}
\ee^{-\ii\RR_\beta(\kk_2-\kk')}
\sum\limits_{\RR_\gamma}
\ee^{\ii\RR_\gamma(\kk_3-\kk')}
\sum\limits_{\RR_\delta}
	\ee^{-\ii\RR_\delta(\kk_4-\kk)}, \quad \bn_i = (n_i, {\bf k}_i).
\end{split}
\end{equation}
\end{widetext}
This expression simplifies to Eq. (\ref{eq:chic}) after using\cite{czycholl2004}
\begin{equation}\label{eq:Rsum}
\sum\limits_{\RR}\ee^{\ii\kk\RR}=\sum\limits_{\GG}\delta_{\kk\GG}
\end{equation}
and the periodicity condition\cite{RevModPhys.84.1419}
\begin{equation}\label{eq:periodicity}
T_{n\alpha}^{(\kk)}=T_{n\alpha}^{(\kk+\GG)}
\end{equation}

\section{Leverage score}\label{app:Leverage}
The matrix $P$ defined in Eq. (\ref{eq:projector}) is a projector and thus
idempotent, i.e. $P^\dagger P = P$. As a consequence the eigenvalues are
discrete and either 0 or 1. This enables direct identification of the singular
eigenvectors. The left singular vectors $U$ correspond to the selected row
vectors of the Wannier projection matrix $T$, while the right singular vectors
$V$ correspond to selected column vectors of $T^\dagger$.  This structure has an
important consequence when combining Eq.  (\ref{eq:leverage}) with Eq.
(\ref{eq:projector}). The leverage scores appear directly as the diagonal
elements in $P$. 
\FloatBarrier

\section*{References}
\bibliography{master}
\end{document}